\documentclass[a4paper,titlepage,thmsb,amsfont]{article}
\usepackage{amsmath}
\usepackage{sw20dft1}
\usepackage{epsfig}
\usepackage{graphicx}
\usepackage{amsfonts}
\usepackage{amssymb}
\begin{document}

\title{Sequential Monte Carlo Samplers}
\author{$\dagger$Pierre Del Moral - $\ddagger$Arnaud Doucet\\$\dagger$CNRS-UMR C55830, Universit\'{e} Paul Sabatier,\\Laboratoire de Statistiques et Probabilit\'{e}s,\\31062 Toulouse, France.\\$\ddagger$Department of Engineering, Cambridge University,\\Trumpington Street, Cambridge CB2 1PZ, UK.\\Email: \texttt{delmoral@cict.fr} - \texttt{ad2@eng.cam.ac.uk}}
\date{CUED-F-INFENG TR. 443 (December 2002)}
\date{CUED-F-INFENG TR. 443 (December 2002)}
\maketitle
\begin{abstract}
\end{abstract}
\begin{abstract}
\end{abstract}

\section{\textbf{Introduction}}

Assume we want to sample from a sequence of probability distributions
$\left\{  \pi_{n}\right\}  _{n\in\mathcal{N}}$ defined on a common measurable
space $E$ where $\mathcal{N}=\left\{  0,1,\ldots,p\right\}  $ or
$\mathcal{N}=\mathbb{N}$. As a special case, one can set $\pi_{n}=\pi$ for all
$n\in\mathcal{N}$. Alternatively the distribution can vary across
$\mathcal{N}$. Similarly to simulated annealing, one could be interested in
the sequence of distributions $\pi_{n}\left(  dx\right)  \propto\pi
^{\gamma_{n}}\left(  x\right)  dx$ for an increasing schedule $\left\{
\gamma_{n}\right\}  _{n\in\mathcal{N}}$ so as to maximize $\pi$ or $\pi_{n}$
could be the posterior distribution of a parameter given the data collected
till time $n$. In this paper, we are interested in sampling this sequence of
distributions \emph{sequentially}; that is first sampling from $\pi_{0}$ then
$\pi_{1}$ and so on. We will further on refer to $n$ as the time index.

The tools favoured by statisticians to achieve this are Markov Chain Monte
Carlo (MCMC) methods; see for example (Robert and Casella, 1999). To sample
from $\pi_{n}$, MCMC\ methods consists of building an ergodic Markov kernel
$K_{n}$ with invariant distribution $\pi_{n}$ using Metropolis-Hastings (MH)
steps, Gibbs steps etc. MCMC have been successfully used in many applications
in statistics and physics. When the distribution to sample is multimodal,
MCMC\ samplers can be easily stucked in one mode. A standard approach to
improve mixing consists of using interacting parallel MCMC/tempering
mechanisms where one runs a MCMC\ chain on an extended space $E^{N}$ with a
specified joint\ invariant distribution admitting $\pi_{n}$ as a marginal
(Geyer and Thompson, 1995). However, MCMC are not well adapted to sequential
simulation. At index $n$, one needs to wait for the Markov chain with kernel
$K_{n}$ to reach its stationary distribution $\pi_{n}$.

We propose here a different approach to sample from $\left\{  \pi_{n}\right\}
_{n\in\mathcal{N}}$. Our approach is based on Sequential Monte Carlo (SMC)
methods (Doucet \emph{et al.}, 2001; Liu, 2001). Henceforth the resulting
algorithms will be called SMC\ samplers. SMC methods have been recently
studied and used extensively in the context of sequential Bayesian inference
and physics (Doucet \emph{et al.}, 2001; Iba, 2001; Liu, 2001). At a given
time $n$, the basic idea is to obtain a large collection of $N$ ($N\gg1$)
random samples $\left\{  X_{n}^{\left(  i\right)  }\right\}  _{i=1,\ldots,N}$
named particles whose marginal distribution is asymptotically ($N\rightarrow
\infty$) equal to $\pi_{n}$. These particles are carried forward over time
using a combination of Sequential Importance Sampling (SIS) and resampling
ideas. This approach is very different from parallel MCMC\ algorithms where
one builds a Markov kernel with a specified joint invariant distribution on
$E^{N}$.

Standard SMC algorithms available in the literature do not apply to our
problem. Indeed, these algorithms deal with the case where each target
distribution of interest $\pi_{n}$ is defined on $E_{n}$ with $E_{n-1}\subset
E_{n}.$ In (Chopin, 2002), a SMC\ algorithm is proposed to deal with the case
$E_{n}=E$. However, this approach restricts severely the way particles can
explore the space. The idea in this paper is different and consists of
building an artificial sequence of distributions $\left\{  \widetilde{\pi}%
_{n}\right\}  _{n\in\mathcal{N}}$ defined on $E_{n}=E^{n+1}$ with
$\widetilde{\pi}_{n}$ admitting a marginal $\pi_{n}$. We are then back to the
standard SMC\ framework. More precisely, $\widetilde{\pi}_{n}$ is defined on
$E^{n+1}$ by $\widetilde{\pi}_{0}\left(  dx_{0}\right)  =\pi_{0}\left(
dx_{0}\right)  $ and
\begin{equation}
\widetilde{\pi}_{n}\left(  d\left(  x_{0},\ldots,x_{n}\right)  \right)
=\pi_{n}\left(  dx_{n}\right)  \prod\limits_{k=1}^{n}L_{k}\left(
x_{k},dx_{k-1}\right) \label{eq:sequencedistributions}%
\end{equation}
where $\left\{  L_{n}\right\}  _{n\in\mathcal{N\backslash}\left\{  0\right\}
} $ is a sequence of auxiliary Markov transition kernels.

Our approach has some connections with Annealed Importance Sampling (AIS)
(Neal, 2001) and the algorithm recently proposed in (Capp\'{e} \emph{et al.},
2002) which are detailed in Section 2. However, the framework we present here
is more general and allows to derive a whole class of principled integration
and genetic-type optimization algorithms based on interacting particle
systems. Similarly to MCMC, the efficiency of the algorithms is dependent on
the target distributions, the proposal and auxiliary kernels. Nevertheless,
generally speaking, one can expect that SMC\ samplers will outperform MCMC
when the distributions to sample are multimodal with well-separated modes.
Moreover SMC\ samplers can be used for sequential Bayesian inference problems
with static parameters like those addressed by Chopin (2002).

This paper focuses on the algorithmic aspects of SMC\ samplers. However, it is
worth noting that our algorithms can be interpreted as an interacting particle
approximating model of a nonlinear Feynman-Kac flow in distribution space.
Under additional assumptions, we provide a nonlinear Markov interpretation of
the measure-valued dynamic system associated to the flow $\left\{  \pi
_{n}\right\}  _{n\in\mathcal{N}}$. We show that this interpretation is a
natural nonlinear version of the MH algorithm. Many convergence results are
available for Feynman-Kac flows and their interacting particle approximations
(Del Moral and Miclo, 2000; Del Moral and Miclo, 2001) and, consequently, for
SMC\ samplers. However, the Feynman-Kac flow associated to SMC\ samplers is
such that many known estimates on the asymptotic behaviour of these
interacting processes can be greatly improved. Several of these results can be
found in (Del Moral and Doucet, 2003).

The rest of the paper is organized as follows. In Section 2, we review a
generic SMC algorithm to sample from the sequence of distributions
(\ref{eq:sequencedistributions}). Various settings for this algorithm are
presented, some extensions and the connections with previous work are
outlined. Section 3 describes the distribution flow associated to our
interacting particle approximating model and also presents an original
nonlinear Markovian interpretation of this flow. Section 4 applies this class
of algorithms to a nonlinear regression problem. Finally, we discuss briefly a
few open methodological and theoretical problems in Section 5.

\section{Sequential Monte Carlo Sampling\label{sec:SMCalgo}}

\subsection{A Generic Algorithm}

We describe here a generic SMC\ algorithm to sample from the sequence of
distributions $\left\{  \pi_{n}\right\}  _{n\in\mathcal{N}}$ defined in
(\ref{eq:sequencedistributions}) based on a Sampling Importance Resampling
strategy; see (Doucet \emph{et al.}, 2001) for a booklength survey of the
SMC\ literature. Alternative SMC algorithms such as the Auxiliary Particle
method of Pitt and Shephard (1999) could also be used.

Further on we will use the notation $X_{0:k}$ to denote $\left(  X_{0}%
,\ldots,X_{k}\right)  $. At time $n-1$, assume a set of particles $\left\{
X_{0:n-1}^{\left(  i\right)  }\right\}  $ ($i=1,\ldots N$) distributed
approximately according to $\widetilde{\pi}_{n-1}$ is available, i.e. the
empirical measure
\[
\widehat{\widetilde{\pi}}_{n-1}\left(  dx_{0:n-1}\right)  =\frac{1}{N}%
\sum_{i=1}^{N}\delta_{X_{0:n-1}^{\left(  i\right)  }}\left(  dx_{0:n-1}%
\right)
\]
is an approximation of $\widetilde{\pi}_{n-1}.$ At time $n$, we extend the
path of each particle according to a Markov\footnote{The Markov assumption
could be relaxed.} kernel $M_{n}\left(  x_{n-1},dx_{n}\right)  $. The
resulting path is thus approximately distributed according to $\widetilde{\pi
}_{n-1}\left(  dx_{0:n-1}\right)  M_{n}\left(  x_{n-1},dx_{n}\right)  $.
Importance sampling can then be used to correct for the discrepancy between
the sampling distribution and $\widetilde{\pi}_{n}\left(  dx_{0:n}\right)  $,
with the importance weight satisfying
\begin{align}
G_{n}\left(  x_{n-1},x_{n}\right)   & =\frac{\widetilde{\pi}_{n}\left(
dx_{0:n}\right)  }{\widetilde{\pi}_{n-1}\left(  dx_{0:n-1}\right)
M_{n}\left(  x_{n-1},dx_{n}\right)  }\label{eq:importanceweight}\\
& =\frac{\pi_{n}\left(  dx_{n}\right)  L_{n}\left(  x_{n},dx_{n-1}\right)
}{\pi_{n-1}\left(  dx_{n-1}\right)  M_{n}\left(  x_{n-1},dx_{n}\right)
}\nonumber
\end{align}
and being assumed well-defined.\ Finally, the particles are resampled
according to their importance weights; particles with low weights are
discarded whereas particles with high weights are multiplied. The resampled
particles are given an equal weight. To sum up, the algorithm proceeds as
follows.\vspace{-0.5cm}%

%TCIMACRO{\TeXButton{hrulefill}{\noindent
%\hrulefill}}%
%BeginExpansion
\noindent
\hrulefill
%EndExpansion
\vspace{-0.6cm}

\begin{center}
\textbf{Sequential Monte Carlo Sampler}\vspace{-0.5cm}
\end{center}

\vspace{-0.5cm}%
%TCIMACRO{\TeXButton{hrulefill}{\noindent
%\hrulefill}}%
%BeginExpansion
\noindent
\hrulefill
%EndExpansion

\vspace{-0.5cm}\hspace{-0.5cm}\textbf{Initialization;} $n=0$.

\underline{\textit{Sampling step}}

\begin{itemize}
\item \textsf{For }$i=1,...,N$\textsf{, sample }$\widetilde{X}_{0}^{\left(
i\right)  }\sim v_{0}\left(  \cdot\right)  $\textsf{.}

\item \textsf{For }$i=1,...,N$\textsf{, evaluate the normalized weights
}$W_{0}^{\left(  i\right)  }$%
\[
W_{0}^{\left(  i\right)  }\propto G_{0}\left(  \widetilde{X}_{0}^{\left(
i\right)  }\right)  =\frac{\pi_{0}\left(  d\widetilde{X}_{0}^{\left(
i\right)  }\right)  }{v_{0}\left(  d\widetilde{X}_{0}^{\left(  i\right)
}\right)  }\text{, }\sum_{i=1}^{N}W_{0}^{\left(  i\right)  }=1.
\]
\end{itemize}

\underline{\textit{Resampling step}}

\begin{itemize}
\item \textsf{Multiply/Discard particles }$\left\{  \widetilde{X}_{0}^{\left(
i\right)  }\right\}  $\textsf{\ with respect to high/low weights }$\left\{
W_{0}^{\left(  i\right)  }\right\}  $ \textsf{to obtain }$N$%
\textsf{\ particles }$\left\{  X_{0}^{\left(  i\right)  }\right\}  $.
\end{itemize}

\hspace{-0.5cm}\textbf{Iteration} $n;$ $n\in\mathcal{N\backslash}\left\{
0\right\}  $.

\underline{\textit{Sampling step}}

\begin{itemize}
\item \textsf{For }$i=1,...,N$\textsf{, set }$\widetilde{X}_{0:n-1}^{\left(
i\right)  }=X_{0:n-1}^{\left(  i\right)  }$ \textsf{and sample }$\widetilde
{X}_{n}^{\left(  i\right)  }\sim M_{n}\left(  \widetilde{X}_{n-1}^{\left(
i\right)  },\cdot\right)  $\textsf{.}

\item \textsf{For }$i=1,...,N$\textsf{, evaluate the normalized weights
}$W_{n}^{\left(  i\right)  }$%
\[
W_{n}^{\left(  i\right)  }\propto G_{n}\left(  \widetilde{X}_{n-1}^{\left(
i\right)  },\widetilde{X}_{n}^{\left(  i\right)  }\right)  \text{, }\sum
_{i=1}^{N}W_{n}^{\left(  i\right)  }=1.
\]
\end{itemize}

\underline{\textit{Resampling step}}

\begin{itemize}
\item \textsf{Multiply/Discard particles }$\left\{  \widetilde{X}%
_{0:n}^{\left(  i\right)  }\right\}  $\textsf{\ with respect to high/low
weights }$\left\{  W_{n}^{\left(  i\right)  }\right\}  $ \textsf{to obtain
}$N$\textsf{\ particles }$\left\{  X_{0:n}^{\left(  i\right)  }\right\}
$.\vspace{-0.8cm}
\end{itemize}%

%TCIMACRO{\TeXButton{hrulefill}{\noindent
%\hrulefill}}%
%BeginExpansion
\noindent
\hrulefill
%EndExpansion

In this algorithm, $v_{0}$ is the initial importance distribution. The
resampling step can be done using a standard procedure such as multinomial
resampling (Gordon \emph{et al.}, 1993), stratified resampling (Kitagawa,
1996) or minimum entropy resampling (Crisan, 2001). All these resampling
schemes are unbiased; that is the number of times $N_{i}$ the particle
$\widetilde{X}_{0:n}^{\left(  i\right)  }$ is copied satisfies $E\left(
N_{i}\right)  =NW_{n}^{\left(  i\right)  }$. MCMC\ steps with invariant
distribution $\widetilde{\pi}_{n}$ can also be included after the resampling
step (Gilks and Berzuini, 1999).

The complexity of this algorithm is in $O\left(  N\right)  $ and it can be
parallelized easily. In practice, the memory requirements are in $O\left(
N\right)  $ too and do not increase over time as one does not need to keep in
memory at time $n$ the whole paths $\left\{  X_{0:n}^{\left(  i\right)
}\right\}  $ but only $\left\{  X_{n}^{\left(  i\right)  }\right\}  $.

The algorithm can be interpreted as an adaptive importance sampling strategy.
Initially, $v_{0}$ is used and the particles with the highest importance
weights are multiplied whereas the ones with small weights are discarded. At
time $n$, new ``candidate'' particles are sampled according to a proposal
distribution kernel $M_{n}$. If $M_{n}$ is a random walk, then the new
particles can be interpreted as a local exploration of the distribution. The
crucial point is that these candidates are weighted by
(\ref{eq:importanceweight}) so as to ensure that after the resampling step
their distribution is approximately $\widetilde{\pi}_{n}$. The introduction of
the auxiliary kernel $L_{n}$ allows the use of importance sampling without
having to compute the marginal distribution $\int\pi_{n-1}\left(  du\right)
M_{n}\left(  u,dx\right)  $ of the particles $\left\{  \widetilde{X}%
_{n}^{\left(  i\right)  }\right\}  $. Indeed, this marginal importance
distribution does not typically admit an analytical expression except when
$M_{n}$ is a MCMC\ kernel of invariant distribution $\pi_{n-1}$. A similar
idea is the basis of the conditional Monte Carlo method described by
Hammersley (1956).

\subsection{Particle Estimates}

At time $n$, we have the following empirical approximations of $\pi_{n}$
before the resampling step
\[
\widehat{\pi}_{n,1}\left(  dx\right)  =\sum_{i=1}^{N}W_{n}^{\left(  i\right)
}\delta_{\widetilde{X}_{n}^{\left(  i\right)  }}\left(  dx\right)  .
\]
and after the resampling step it is equal to
\[
\widehat{\pi}_{n,2}\left(  dx\right)  =\frac{1}{N}\sum_{i=1}^{N}\delta
_{X_{n}^{\left(  i\right)  }}\left(  dx\right)  .
\]
For any measure $\mu$ and function $f$, we will denote $\mu\left(  f\right)
=\int f\left(  x\right)  \mu\left(  dx\right)  $. An estimate of $\pi
_{n}\left(  f\right)  $ is given by
\[
\int f\left(  x\right)  \widehat{\pi}_{n,1}\left(  dx\right)  =\sum_{i=1}%
^{N}W_{n}^{\left(  i\right)  }f\left(  \widetilde{X}_{n}^{\left(  i\right)
}\right)  .
\]
or alternatively $\int f\left(  x\right)  \widehat{\pi}_{n,2}\left(
dx\right)  =\frac{1}{N}\sum_{i=1}^{N}f\left(  X_{n}^{\left(  i\right)
}\right)  $ which has higher variance. If $\pi_{n}=\pi$, then the following
estimate can be also used
\begin{equation}
\frac{1}{n}\sum_{k=1}^{n}\sum_{i=1}^{N}W_{k}^{\left(  i\right)  }f\left(
\widetilde{X}_{k}^{\left(  i\right)  }\right)  .\label{eq:ergodicaverage}%
\end{equation}
Though the particles are statistically dependent, one can show under
assumptions given in (Del Moral and Miclo, 2000) that this estimate is
consistent as $N\rightarrow\infty$.

The algorithm described above can also be used to compute the ratio of
normalizing constants. Indeed, typically the sequence of distributions
$\pi_{n}\left(  dx\right)  $ is only known up to a normalizing constant, i.e.
say $\pi_{n}\left(  dx\right)  \propto f_{n}\left(  x\right)  dx$. In this
case, the unnormalized importance weights one computes are equal to
\[
\widetilde{W}_{n}^{\left(  i\right)  }=\frac{f_{n}\left(  \widetilde{X}%
_{n}^{\left(  i\right)  }\right)  L_{n}\left(  \widetilde{X}_{n}^{\left(
i\right)  },d\widetilde{X}_{n-1}^{\left(  i\right)  }\right)  }{f_{n-1}\left(
\widetilde{X}_{n-1}^{\left(  i\right)  }\right)  M_{n}\left(  \widetilde
{X}_{n-1}^{\left(  i\right)  },d\widetilde{X}_{n}^{\left(  i\right)  }\right)
}\propto W_{n}^{\left(  i\right)  }%
\]
at time $n$. It is possible to obtain an estimate of the ratio of the
normalizing constants
\[
\frac{Z_{n}}{Z_{n-1}}=\frac{\int f_{n}\left(  x\right)  dx}{\int
f_{n-1}\left(  x\right)  dx}%
\]
using
\begin{equation}
\widehat{\frac{Z_{n}}{Z_{n-1}}}=\frac{1}{N}\sum_{i=1}^{N}\widetilde{W}%
_{n}^{\left(  i\right)  }.\label{eq:rationormestimate}%
\end{equation}
Thus an estimate of $\log\left(  Z_{n}/Z_{0}\right)  $ is given by
\[
\log\left(  \widehat{\frac{Z_{n}}{Z_{0}}}\right)  =\sum_{k=1}^{n}\log\left(
\widehat{\frac{Z_{k}}{Z_{k-1}}}\right)  =\sum_{k=1}^{n}\log\left(  \sum
_{i=1}^{N}\widetilde{W}_{k}^{\left(  i\right)  }\right)  -n\log N.
\]
If the resampling scheme used is unbiased, then (\ref{eq:rationormestimate})
is also unbiased (Del Moral and Miclo, 2000).

\subsection{Algorithm Settings}

The algorithm presented in the previous subsection is very general. There are
many potential choices for $\left\{  \pi_{n},M_{n},L_{n}\right\}
_{n\in\mathcal{N}}$ leading to various integration and optimization algorithms.

\emph{Homogeneous sequences.} A simple choice consists of setting $\pi_{n}%
=\pi,$ $M_{n}=M$ and $L_{n}=L$. In this case, the importance weight
(\ref{eq:importanceweight}) is the following generalized MH ratio
\begin{equation}
G\left(  x,x^{\prime}\right)  =\frac{\pi\left(  dx^{\prime}\right)  L\left(
x^{\prime},dx\right)  }{\pi\left(  dx\right)  M\left(  x,dx^{\prime}\right)
};\label{eq:ratiometropolis}%
\end{equation}
the standard MH ratio corresponds to $M=L$. In this simple case, the particles
evolve independently according to a proposal distribution $M$, their
generalized MH\ ratio is computed and normalized. The particles are then
multiplied or discarded with respect to the value of their normalized MH ratio.

\emph{Sequence of distributions} $\pi_{n}$. It might be of interest to
consider non homogeneous sequence of distributions either to move ``smoothly''
from $\pi_{0}=v_{0}$ to a target distribution $\pi$ through a sequence of
intermediate distributions or for the sake of optimization. In the case of
integration as suggested by Neal (2001), one can select
\[
\pi_{n}\left(  dx\right)  \propto\pi^{\gamma_{n}}\left(  x\right)  \pi
_{0}^{1-\gamma_{n}}\left(  x\right)  dx
\]
with $\mathcal{N}=\left\{  0,\ldots,p\right\}  $,$\mathcal{\ }\gamma_{0}=0$
and $\gamma_{p}=1$. For the case of optimization, one can select
\[
\pi_{n}\left(  dx\right)  \propto\pi^{\gamma_{n}}\left(  x\right)  dx
\]
where $\mathcal{N}=\mathbb{N}$, $\left\{  \gamma_{n}\right\}  _{n\geq0}$ is an
increasing sequence such that $\gamma_{n}\rightarrow\infty$. In this case, the
resulting algorithm is a genetic algorithm where the sampling step is the
``mutation'' step and the resampling step is the selection step (Goldberg,
1989). However, there is a significant difference with standard genetic
algorithms as we know the asymptotic ($N\rightarrow\infty$) distribution\ of
the particles. This makes the analysis of the resulting algorithm easier than
in cases where this distribution is unknown such as in (Del Moral and Miclo,
2003). Convergence properties of the algorithm are currently under study.

Finally, another application of this algorithm consists of estimating the
sequence of posterior distributions $\pi_{n}\left(  dx\right)  =\pi_{n}\left(
\left.  dx\right|  y_{1,\ldots},y_{n}\right)  $ where $y_{n}$ is an
observation available at time $n$. As briefly discussed in the introduction,
SMC\ algorithms have been recently proposed in this framework by Chopin (2002)
but the\ SIS framework used is somehow restricted: it only allows $M_{n}$ to
be a MCMC\ kernel of invariant distribution $\pi_{n-1}$.

\emph{Sequence of proposal kernels} $M_{n}$ \emph{and} \emph{auxiliary kernels
}$L_{n}$. Any couple of kernels can be used as long as the ratio
(\ref{eq:ratiometropolis}) is well defined. However, one can only expect good
properties of the algorithm if this ratio admits a reasonable variance and
also if $L_{n}$ is mixing. Indeed, loosely speaking, the faster $L_{n}$ mixes,
the faster the SMC algorithm forgets Monte Carlo errors (Del Moral and Doucet, 2003).

In SMC\ algorithms (Doucet \emph{et al.}, 2001), it is known that the
importance sampling distribution minimizing the conditional variance of the
weights at time $n$, i.e. $\left\{  X_{0:n-1}^{\left(  i\right)  }\right\}  $
fixed, is given by
\begin{equation}
M_{n}\left(  x,dx^{\prime}\right)  =\frac{\pi_{n}\left(  dx^{\prime}\right)
L_{n}\left(  x^{\prime},dx\right)  }{\int\pi_{n}\left(  du\right)
L_{n}\left(  u,dx\right)  }.\label{eq:optimalsampling}%
\end{equation}
In this case, the importance weight $G_{n}\left(  x,x^{\prime}\right)  $ is
given by
\begin{equation}
G_{n}\left(  x\right)  =\frac{\int\pi_{n}\left(  du\right)  L_{n}\left(
u,dx\right)  }{\pi_{n-1}\left(  dx\right)  }\label{eq:optimalweight}%
\end{equation}
and is independent of $x^{\prime}$. This allows the resampling step to be
performed before the sampling step.

In standard applications of SMC algorithms, the kernel $L_{n}$ is usually
given by the problem at hand whereas in our setup this kernel is arbitrary and
can be optimized for a given proposal distribution $M_{n}$. One can
alternatively select the kernel $L_{n}$ so as to be able to compute
(\ref{eq:optimalweight}); e.g. a MCMC\ kernel of invariant distribution
$\pi_{n}$, and then sample the particles according to
(\ref{eq:optimalsampling}).

For a fixed $M_{n}$, an alternative natural choice\footnote{thanks to C.
Andrieu} consists of choosing
\begin{equation}
L_{n}\left(  x,dx^{\prime}\right)  =\frac{\pi_{n-1}\left(  dx^{\prime}\right)
M_{n}\left(  x^{\prime},dx\right)  }{\int\pi_{n-1}\left(  du\right)
M_{n}\left(  u,dx\right)  }.\label{eq:optimalsampling2}%
\end{equation}
In this case, the associated importance weight $G_{n}\left(  x,x^{\prime
}\right)  $ is given by
\begin{equation}
G_{n}\left(  x^{\prime}\right)  =\frac{\pi_{n}\left(  dx^{\prime}\right)
}{\int\pi_{n-1}\left(  du\right)  M_{n}\left(  u,dx^{\prime}\right)
}.\label{eq:optimalweight2}%
\end{equation}
If $M_{n}$ is a MCMC kernel of invariant distribution $\pi_{n-1}$, then the
weight (\ref{eq:optimalweight2}) can be computed easily. If not, numerical
integration using the current set of particles can be used to approximate it
but the resulting algorithms would be of complexity $O\left(  N^{2}\right)  $.

\subsection{Connections to previous work and Extensions}

\emph{Connections to previous work.} AIS is a method proposed recently by Neal
(2001). Reversing the time index in (Neal, 2001) to be consistent with our
notation, AIS corresponds to the case where one considers a finite sequence of
distributions, $M_{n}$ is a MCMC\ kernel of invariant distribution $\pi_{n}$
and
\begin{equation}
L_{n}\left(  x,dx^{\prime}\right)  =M_{n}\left(  x^{\prime},dx\right)
\frac{\pi_{n}\left(  dx^{\prime}\right)  }{\pi_{n}\left(  dx\right)
}.\label{eq:nealchoice}%
\end{equation}
For a given $M_{n}$, one can check that this choice of $L_{n}$ ensures that
(\ref{eq:optimalsampling})\ is satisfied. In this case, one obtains by
combining (\ref{eq:ratiometropolis})\ and (\ref{eq:nealchoice})
\[
G_{n}\left(  x,x^{\prime}\right)  =G_{n}\left(  x\right)  =\frac{\pi
_{n}\left(  dx\right)  }{\pi_{n-1}\left(  dx\right)  }.
\]
The resampling step is not included in the AIS\ algorithm. In our framework,
we point out that this is a crucial step to include to make the method
efficient as established theoretically in (Del\ Moral and Miclo, 2000) and
practically in our simulations. Otherwise the method is just a special
instance of SIS and collapses\ if $n$ is too large. In (Godsill and Clapp,
2001), the authors used the AIS\ algorithm in combination with resampling in
the context of optimal filtering.

A more recent work (Capp\'{e} \emph{et al.}, 2002) contemporary of (Del Moral
and Doucet, 2003) and developed independently is another special case of our
framework. In (Capp\'{e} \emph{et al.}, 2002), the authors consider the
homogeneous case. Their algorithm corresponds to the case where $M$ is an MCMC
kernel of invariant distribution $\pi$ (namely a Gibbs sampler) and $L\left(
x,dx^{\prime}\right)  =\pi\left(  dx^{\prime}\right)  $, it follows that
\[
G\left(  x,x^{\prime}\right)  =\frac{\pi\left(  dx^{\prime}\right)  }{M\left(
x,dx^{\prime}\right)  }.
\]
This particular case has limited applications as $G\left(  x,x^{\prime
}\right)  $ would not be defined in most applications; e.g. $\pi\left(
dx^{\prime}\right)  =\pi\left(  x^{\prime}\right)  dx^{\prime}$ and $M$ is an
MH kernel.

\emph{Extensions.} The algorithm described\ in this section must be
interpreted as the basic element of more complex algorithms. It is what the
MH\ algorithm is to MCMC. For complex MCMC\ problems, one typically uses a
combination of MH steps where the $n_{x}$ components of $x$ say $\left(
x_{1},\ldots,x_{n_{x}}\right)  $ are updated by subblocks (Robert and Casella,
1999). Similarly, to sample from high dimensional distributions, a practical
SMC\ sampler can update the components of $x$ via subblocks. There are also
numerous potential extensions:

\begin{itemize}
\item  It is straightforward to develop a version of the algorithm so as to
sample distributions defined on an union of subspaces of different dimensions.
However, contrary to reversible jump MCMC algorithms (Green, 1995), no
reversibility condition is needed.

\item  As suggested in (Crisan and Doucet, 2000), one can use a proposal
kernel whose parameters are a function of the whole set of current particles.
This allows the algorithm to automatically scale the proposal distribution
based on the previous importance weights.

\item  In the general case, the sequence of probability distributions
$\left\{  \pi_{n}\right\}  _{n\in\mathcal{N}}$ of interest is such that
$\pi_{n}$ is defined on $E_{n}$ and not on $E$. We can generalize the
algorithm described in this section to this case. We introduce an auxiliary
kernel $L_{n}$ from $E_{n}$ to $E_{n-1}$ and a proposal kernel $M_{n}$ from
$E_{n-1}$ to $E_{n}$. At time $n-1$, $N$ particles $\left\{  X_{n-1}^{\left(
i\right)  }\right\}  $ approximately distributed according to $\pi_{n-1}$ are
available. At time $N$ new particles $\left\{  X_{n}^{\left(  i\right)
}\right\}  $ are sampled according to $X_{n}^{\left(  i\right)  }\sim
M_{n}\left(  X_{n-1}^{\left(  i\right)  },\cdot\right)  $ and the following
importance weights are computed
\[
W_{n}^{\left(  i\right)  }\propto\frac{\pi_{n}\left(  dX_{n}^{\left(
i\right)  }\right)  L_{n}\left(  X_{n}^{\left(  i\right)  },dX_{n-1}^{\left(
i\right)  }\right)  }{\pi_{n-1}\left(  dX_{n-1}^{\left(  i\right)  }\right)
M_{n}\left(  X_{n-1}^{\left(  i\right)  },dX_{n}^{\left(  i\right)  }\right)
}.
\]
Then the particles are resampled.
\end{itemize}

\section{Feynman-Kac Representation and Particle
Interpretations\label{sec:feynmankac}}

In this Section, we show that the algorithm presented in Section
\ref{sec:SMCalgo} corresponds to an interacting particle approximation model
of a nonlinear Feynman-Kac flow in distribution space. We provide an
alternative nonlinear Markovian representation of this flow and its
interacting particle approximation. Here the Feynman-Kac flow corresponds to
the special case where the so-called potential function is given by the
generalized Metropolis ratio (\ref{eq:ratiometropolis}). The abstract
description and the analysis of general Feynman-Kac flows and their particle
approximations have been investigated in several recent research articles.
Many asymptotic ($n\rightarrow\infty$ and/or $N\rightarrow\infty$) results are
available in this field including empirical process convergence, central limit
theorems, large deviation principles as well as increasing propagation of
chaos estimates and uniform convergence estimates with respect to the time
parameter; all of which can be used for SMC\ samplers. The interested reader
is referred to the survey article (Del Moral and Miclo, 2000) and the more
recent studies (Del Moral and Miclo, 2001; Del Moral and Miclo, 2003). As
mentioned in the introduction, the particular choice of the potential function
(\ref{eq:ratiometropolis}) simplifies the analysis and many known estimates on
the asymptotic behaviour of these interacting processes can be greatly
improved. Several of these results can be found in (Del Moral and Doucet, 2003).

\subsection{Feynman-Kac Representation}

Define the following distributions on $E^{2}=E\times E$
\[
\left(  \pi_{n}\times L_{n}\right)  \left(  d\left(  x,x^{\prime}\right)
\right)  =\pi_{n}\left(  dx^{\prime}\right)  L_{n}\left(  x^{\prime
},dx\right)  .
\]
Using (\ref{eq:sequencedistributions}) and (\ref{eq:importanceweight}), it is
clear that the sequence of distributions $\left\{  \pi_{n}\times
L_{n}\right\}  _{n\in\mathcal{N}}$ (with the convention $\pi_{0}\times
L_{0}=\pi_{0}\times\pi_{0}$) admits the following so-called Feynman-Kac
representation
\[
\left(  \pi_{n}\times L_{n}\right)  \left(  f\right)  =\lambda_{n}\left(
f\right)  /\lambda_{n}\left(  1\right)  ,
\]
with
\[
\lambda_{n}\left(  f\right)  =\mathbb{E}_{v_{0},\left\{  M_{k}\right\}
}\left(  f\left(  X_{n-1},X_{n}\right)  G_{0}\left(  X_{0}\right)
\prod\limits_{k=1}^{n}G_{k}\left(  X_{k-1},X_{k}\right)  \right)  ,
\]
where $\mathbb{E}_{v_{0},\left\{  M_{k}\right\}  }$ denotes the expectation
with respect to
\[
v_{0}\left(  dx_{0}\right)  \prod\limits_{k=1}^{n}M_{k}\left(  x_{k-1}%
,dx_{k}\right)  .
\]
This representation is at the core of the results given in (Del Moral and
Miclo, 2000). We give now two ``operator-like'' interpretations of the
sequence $\left\{  \left(  \pi_{n}\times L_{n}\right)  \right\}
_{n\in\mathcal{N}}$. For a measure $\mu$ and a Markov kernel $K$, we use the
standard notation
\[
\mu K\left(  A\right)  =\int_{A}\mu\left(  dz\right)  K\left(  z,dz^{\prime
}\right)  .
\]
Let $\mathcal{P}\left(  E^{2}\right)  $ be the set of probability measures on
$E^{2}$. The mapping $\Psi_{n}:\mathcal{P}\left(  E^{2}\right)  \rightarrow
\mathcal{P}\left(  E^{2}\right)  $ is defined as
\begin{equation}
\pi_{n}\times L_{n}=\Psi_{n}\left(  \left(  \pi_{n-1}\times L_{n-1}\right)
\widetilde{M}_{n}\right) \label{eq:standardrep0}%
\end{equation}
where
\begin{equation}
\Psi_{n}\left(  \mu\right)  \left(  d\left(  u,v\right)  \right)  =\frac
{\mu\left(  d\left(  u,v\right)  \right)  G_{n}\left(  u,v\right)  }%
{\mu\left(  G_{n}\right)  }\label{eq:standardrep}%
\end{equation}
and $\widetilde{M}_{n}$ is a Markov kernel on $E^{2}$ defined as
\[
\widetilde{M}_{n}\left(  \left(  u,v\right)  ,d\left(  u^{\prime},v^{\prime
}\right)  \right)  =\delta_{v}\left(  du^{\prime}\right)  M_{n}\left(
u^{\prime},dv^{\prime}\right)  .
\]
Assuming that $G_{n}$ can be upper bounded over $E^{2}$, one can easily check
that an alternative representation is given by
\begin{equation}
\left(  \pi_{n}\times L_{n}\right)  =\left(  \pi_{n-1}\times L_{n-1}\right)
\widetilde{M}_{n}S_{_{n},\left(  \pi_{n-1}\times L_{n-1}\right)  \widetilde
{M}_{n}}\label{eq:nonlinearmarkovchain}%
\end{equation}
where
\begin{equation}
S_{n,\mu}\left(  \left(  u,v\right)  ,d\left(  u^{\prime},v^{\prime}\right)
\right)  =\epsilon G_{n}\left(  u,v\right)  \delta_{\left(  u,v\right)
}\left(  d\left(  u^{\prime},v^{\prime}\right)  \right)  +\left(  1-\epsilon
G\left(  u,v\right)  \right)  \Psi_{n}\left(  \mu\right)  \left(  d\left(
u^{\prime},v^{\prime}\right)  \right)  ,\label{eq:selectionoperator}%
\end{equation}
$\epsilon$ being chosen such that $\epsilon G_{n}\left(  u,v\right)  \leq1$
over $E^{2}$.

The kernel $\widetilde{M}_{n}S_{_{n},\left(  \pi_{n-1}\times L_{n-1}\right)
\widetilde{M}_{n}}$ is a so-called nonlinear Markov kernel; i.e. the
transition kernel is dependent not only on the current state but also on its
distribution. A\ generic nonlinear Markov chain $\left\{  Z_{n}\right\}
_{n\geq0}$ satisfies
\[
Z_{n}\sim K_{n,\text{Law}\left(  Z_{n-1}\right)  }\left(  Z_{n-1}%
,\cdot\right)  .
\]
It is typically impossible to simulate a realization from such a Markov chain
as the distribution of the state is not available. However, a particle
approximation of it can be used. Consider a Markov chain $\left\{
Z_{n}\right\}  _{n\geq0}$ taking values in $E^{2}$ with transition kernel
$\widetilde{M}_{n}S_{_{n},\left(  \pi_{n-1}\times L_{n-1}\right)
\widetilde{M}_{n}}$. This kernel can be interpreted as follows. Given
$Z_{n-1}=\left(  U_{n-1},V_{n-1}\right)  \sim\left(  \pi_{n-1}\times
L_{n-1}\right)  $, one first sample a candidate $Z_{n}^{\ast}=\left(
U_{n}^{\ast},V_{n}^{\ast}\right)  =\left(  V_{n-1},V_{n}^{\ast}\right)  $
where $V_{n}^{\ast}\sim M_{n}\left(  U_{n}^{\ast},\cdot\right)  $. With
probability $\epsilon G_{n}\left(  U_{n}^{\ast},V_{n}^{\ast}\right)  $, one
sets $Z_{n}=\left(  U_{n}^{\ast},V_{n}^{\ast}\right)  $, otherwise $Z_{n}%
\sim\Psi_{n}\left(  \left(  \pi_{n-1}\times L_{n-1}\right)  \widetilde{M}%
_{n}\right)  $. By construction, one has $Z_{n}\sim\left(  \pi_{n}\times
L_{n}\right)  $. This algorithm can be interpreted as a nonlinear version of
the MH algorithm. The main difference being that, when a candidate is
rejected, the chain does not stay where it is a new state is proposed
according to $\Psi_{n}\left(  \left(  \pi_{n-1}\times L_{n-1}\right)
\widetilde{M}_{n}\right)  $.

\subsection{Particle Interpretations}

The first particle interpretation of the flow follows (\ref{eq:standardrep0}%
)-(\ref{eq:standardrep}). It corresponds to the standard algorithm which has
been described in Section \ref{sec:SMCalgo}. The second alternative algorithm
corresponds to a particle interpretation of the flow corresponding to
(\ref{eq:nonlinearmarkovchain}). It proceeds as follows.

\vspace{-0.5cm}%
%TCIMACRO{\TeXButton{hrulefill}{\noindent
%\hrulefill}}%
%BeginExpansion
\noindent
\hrulefill
%EndExpansion

\vspace{-0.5cm}\textbf{Iteration} $n;$ $n\in\mathcal{N\backslash}\left\{
0\right\}  $.

\underline{\textit{Sampling step}}

\begin{itemize}
\item \textsf{For }$i=1,...,N$\textsf{, set }$X_{0:n-1}^{\left(  i\right)
}=\widetilde{X}_{0:n-1}^{\left(  i\right)  }$ \textsf{and sample }%
$\widetilde{X}_{n}^{\left(  i\right)  }\sim M_{n}\left(  \widetilde{X}%
_{n-1}^{\left(  i\right)  },\cdot\right)  $\textsf{.}

\item \textsf{For }$i=1,...,N$\textsf{, evaluate the normalized weights
}$W_{n}^{\left(  i\right)  }$%
\begin{equation}
W_{n}^{\left(  i\right)  }\propto G_{n}\left(  \widetilde{X}_{n-1}^{\left(
i\right)  },\widetilde{X}_{n}^{\left(  i\right)  }\right)  \text{, }\sum
_{i=1}^{N}W_{n}^{\left(  i\right)  }=1.
\end{equation}
\end{itemize}

\underline{\textit{Resampling step}}

\begin{itemize}
\item $J=\emptyset.$

\item \textsf{For }$i=1,...,N$\textsf{, with probability }$\epsilon
G_{n}\left(  \widetilde{X}_{n-1}^{\left(  i\right)  },\widetilde{X}%
_{n}^{\left(  i\right)  }\right)  $\textsf{, set }$X_{0:n}^{\left(  i\right)
}=\widetilde{X}_{0:n}^{\left(  i\right)  }$ \textsf{otherwise set }%
$J=J\cup\left\{  i\right\}  .$

\item \textsf{Multiply/Discard particles }$\left\{  \widetilde{X}%
_{0:n}^{\left(  i\right)  }\right\}  $\textsf{\ with respect to high/low
weights }$\left\{  W_{n}^{\left(  i\right)  }\right\}  $ \textsf{to obtain
}$\left\{  X_{0:n}^{\left(  i\right)  }\right\}  _{i\in J}$.\vspace{-0.5cm}
\end{itemize}%

%TCIMACRO{\TeXButton{hrulefill}{\noindent
%\hrulefill}}%
%BeginExpansion
\noindent
\hrulefill
%EndExpansion

\section{Simulation Results}

\subsection{Model}

We consider the following harmonic regression model (Andrieu and Doucet,
1999)
\[
Y=D\left(  \omega\right)  \beta+n,
\]
where $Y=\left(  y_{0},\ldots,y_{m-1}\right)  ^{\mathtt{T}}$, $\beta=\left(
\beta_{1},\ldots,\beta_{2k}\right)  ^{\mathtt{T}},$ $n=\left(  n_{0}%
,\ldots,n_{m-1}\right)  ^{\mathtt{T}}$, $\omega=\left(  \omega_{1}%
,\ldots,\omega_{k}\right)  ^{\mathtt{T}}\in\left(  0,\pi\right)  ^{k}$ and
$D\left(  \omega\right)  $ is a $m\times2k$ matrix where for $i=0,\ldots
,m-1,\;j=1,\ldots,k $ .
\[
\left[  D\left(  \omega\right)  \right]  _{i+1,2j-1}=\cos\left(  \omega
_{j}i\right)  ,\text{ }\left[  D\left(  \omega\right)  \right]  _{i+1,2j}%
=\sin\left(  \omega_{j}i\right)  .
\]
We assume that $\left.  n\right|  \sigma^{2}\sim\mathcal{N}\left(
0,\sigma^{2}I_{p}\right)  $ and we use the following prior $p\left(
\sigma^{2},\beta,\omega\right)  =p\left(  \omega\right)  p\left(  \left.
\beta\right|  \sigma^{2}\right)  p\left(  \sigma^{2}\right)  $ with
\[
\sigma^{2}\sim\mathcal{IG}\left(  \frac{\upsilon_{0}}{2},\frac{\gamma_{0}}%
{2}\right)  ,\text{ }\left.  \beta\right|  \sigma^{2}\sim\mathcal{N}\left(
0,\sigma^{2}\mathbf{\Sigma}_{0}\right)  ,
\]
where $\mathbf{\Sigma}_{0}^{-1}=\delta^{-2}D^{\mathtt{T}}\left(
\omega\right)  D\left(  \omega\right)  $ ($\delta^{2}=25$), $\upsilon
_{0}=\gamma_{0}=1$; $p\left(  \omega\right)  $ is uniform on $\Omega=\left\{
\omega\in\left(  0,\pi\right)  ^{k};0<\omega_{1}<\ldots<\omega_{k}%
<\pi\right\}  $. The posterior density satisfies on $\Omega$
\[
p\left(  \left.  \omega\right|  Y\right)  \propto\left(  \gamma_{0}%
+Y^{\mathtt{T}}PY\right)  ^{-\frac{p+\upsilon_{0}}{2}}%
\]
with
\[%
\begin{tabular}
[c]{l}%
$M^{-1}=\left(  1+\delta^{-2}\right)  D^{\mathtt{T}}\left(  \omega\right)
D\left(  \omega\right)  ,$\\
$m=MD^{\mathtt{T}}\left(  \omega\right)  Y,$\\
$P=I_{p}-D\left(  \omega\right)  MD^{\mathtt{T}}\left(  \omega\right)  .$%
\end{tabular}
\]
We simulate a realization of $m=100$ observations with $k=6$, $\sigma^{2}=5$,
\begin{align*}
\omega & =\left(  0.08,0.13,0.21,0.29,0.35,0.42\right)  ^{\mathtt{T}},\\
\beta & =\left(
1.24,0.00,1.23,0.43,0.67,1.00,1.11,0.39,1.31,0.16,1.28,0.13\right)
^{\mathtt{T}}.
\end{align*}
The posterior density is multimodal with well-separated modes.

\subsection{Algorithms}

To sample from $\pi\left(  \omega\right)  =p\left(  \left.  \omega\right|
Y\right)  ,$ we use an homogeneous SMC\ sampler with $N=1000$ particles where
the $k$ components are updated one-at-a-time using a simple Gaussian random
walk proposal $M$ of standard deviation $\sigma_{RW}$. We select $L$ to be
equal to $M$ and use the stratified resampling procedure. We compare our
algorithm with a MCMC\ algorithm. The MCMC\ algorithm updates the component
one-at-a-time using a MH\ step with the proposal kernel $M$. In both case, the
initial distribution is the uniform distribution on $\Omega.$

We consider the case where $\sigma_{RW}=0.1$. Obviously one could come up with
a better proposal kernel. We want to emphasize here that the SMC\ approach is
more robust to a poor scaling of the proposal. A similar remark was made in
(Capp\'{e} \emph{et al.}, 2002). In Figure 1, we present the marginal
posterior distributions of $\omega_{1}$ and $\omega_{2}$ obtained using the
SMC\ sampler with $100$ iterations. We then run $12000$ iterations of the
MCMC\ algorithm so as the computational complexity to be roughly the same for
the two algorithms. The MCMC\ algorithm is more sensitive to the
initialization. On $50$ realizations of the SMC\ and the MCMC\ algorithm, the
SMC always explores the main mode whereas the MCMC\ algorithm converges
towards it only $36$ times.

We also use an inhomogeneous version of the SMC\ sampler so as to optimize
$p\left(  \left.  \omega\right|  Y\right)  $. In this case the target density
at time $n$ is $\pi_{n}\left(  \omega\right)  \propto p^{\gamma_{n}}\left(
\left.  \omega\right|  Y\right)  $ with $\gamma_{n}=n$ and we use $50$
iterations. We compare this algorithm to a simulated annealing version of the
MH algorithm with $60000$ iterations with $\gamma_{n}=n/1200$. In Table 1, we
display the mean and standard deviations of the log-posterior density of the
posterior mode estimate; the posterior mode estimate being chosen as the
sample generated during the simulation maximizing the posterior density.
Contrary to the simulated annealing algorithm, the SMC algorithm converges
consistently towards the same mode.

\bigskip%

%TCIMACRO{\TeXButton{BeginFigureEPS}{
%\begin{figure}[h!]
%	\leavevmode
%	\begin{center}
%\epsfxsize=10cm \epsfysize=8cm \epsffile{fig1.eps}
%	\end{center}
%}}%
%BeginExpansion
\begin{figure}[h!]
	\leavevmode
	\begin{center}
\epsfxsize=10cm \epsfysize=8cm \epsffile{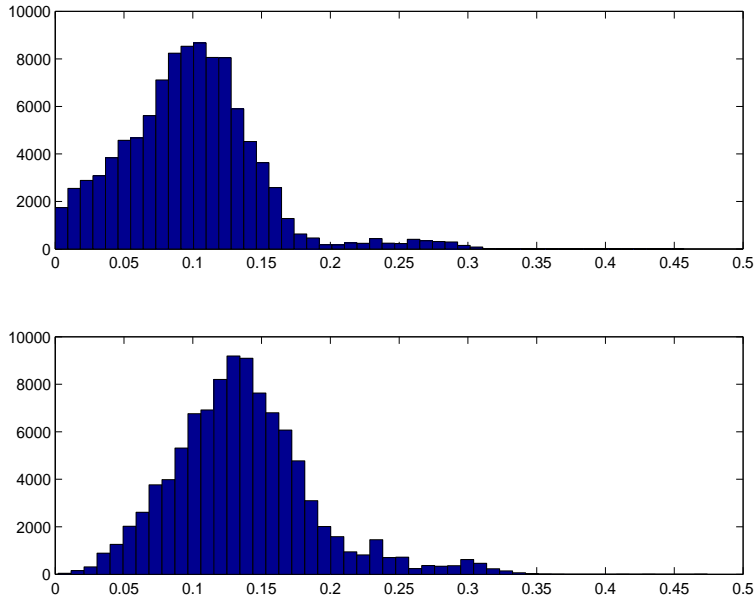}
	\end{center}
%EndExpansion

\caption{Histograms of the simulated values of $\left( \omega_{1},\omega
_{2}\right)
$ using SMC - Estimation of $p\left( \left. \omega_{1}\right| Y\right) $
(top) and $p\left( \left. \omega_{2}\right| Y\right) $ (bottom).}%

%TCIMACRO{\TeXButton{EndFigureEPS}{\end{figure}
%\vspace*{5mm}
%}}%
%BeginExpansion
\end{figure}
\vspace*{5mm}
%EndExpansion
\bigskip

\bigskip

\begin{center}
\bigskip%

\begin{tabular}
[c]{lll}\hline
Algorithm & SMC & MCMC\\\hline
Mean of the log-posterior values & -326.12 & -328.87\\
Standard deviation of the\vspace{-0.5cm} & 0.12 & 1.48\\
log-posterior values &  & \\\hline
\end{tabular}

\bigskip

Table 1: Performance of SMC\ and MCMC\ algorithm obtained over 50 simulations
\end{center}

\section{Discussion}

In this article, we have presented a class of methods to sample from
distributions known up to a normalizing constant. These methods are based on
SMC algorithms. This framework is very general and flexible. Several points
not discussed here are detailed in (Del Moral and Doucet, 2003).

\begin{itemize}
\item  In the homogeneous case, assume that we do not initialize the algorithm
in the stationary regime, i.e. we do not correct for the discrepancy between
$v_{0}$ and $\pi$. This has to be paralleled with MCMC algorithms which are
not initialized in the stationary regime. Under regularity assumptions, it can
be shown that the distribution flow still converges towards the target
distribution $\pi$. Moreover, it converges at a rate only dependent on the
mixing properties of $L$. This is in contrast with the MH\ algorithm whose
rate of convergence is dependent on $\pi$ and $M$.

\item  The algorithm we have presented can be used to simulate a Markov chain
with a fixed terminal point. Indeed, one obtains at time $n$ samples from
(\ref{eq:sequencedistributions}). By setting $L_{0}\left(  x_{1}%
,dx_{0}\right)  =\delta_{x}\left(  dx_{0}\right)  $ and reversing the time
index, one obtains an approximate realization of a Markov process of initial
distribution $\pi_{n}$ at time $0$, transition $\left\{  L_{n\text{ }%
}\right\}  $ and terminal point $x$ at time $n+1$. This has applications in
genetics and physics.
\end{itemize}

There are also several important open methodological and theoretical problems
to study.

\begin{itemize}
\item  Similarly to MCMC\ methods, one needs to carefully design the various
components of the algorithm to get good performance. In particular, it would
be of interest to come up with an automated choice for $L_{n}$ given $M_{n}$.
For the homogeneous case, one could look at minimizing the variance of
(\ref{eq:ergodicaverage}). It involves a tradeoff between the mixing
properties of $L$ and the variance of the importance weights
(\ref{eq:importanceweight}). This point is currently under study.

\item  It would be interesting to weaken the assumptions of the results in
(Del Moral and Miclo, 2000; Del Moral and Doucet, 2003) which mostly only hold
for compact spaces.
\end{itemize}

\section{Acknowledgments}

The authors are extremely grateful to Manuel Davy for his comments and some of
the numerical simulations.\ We are also grateful to Christophe\ Andrieu and
Elena Punskaya for their comments.

\end{document}